\documentclass[cits]{PoS}

\title{Recent Progress in the Understanding of the $r$-Process}

\ShortTitle{Recent Progress in the Understanding of the $r$-Process}

\author{\speaker{Yong-Zhong Qian}%
         \thanks{This work was supported in part by US DOE grant
         DE-FG02-87ER40328.}\\
        School of Physics and Astronomy, University of Minnesota, Minneapolis, MN 55455, USA\\
        E-mail: \email{qian@physics.umn.edu}}

\abstract{A brief overview of the $r$-process is given with an emphasis on
the observational implications for this process. The conditions required for
the major production of the heavy $r$-process elements ($r$-elements)
with mass numbers
$A>130$ are discussed based on a generic astrophysical model where
matter adiabatically expands from a hot and dense initial state. Nucleosynthesis
in the neutrino-driven winds from nascent neutron stars is discussed as a
specific example. Such winds readily produce the elements from Sr to Ag 
with $A\sim 88$--110 through charged-particle reactions in the $\alpha$-process
but appear incapable of making the heavy $r$-elements.
Observations of elemental abundances in metal-poor
stars have provided many valuable insights into the $r$-process.
They have demonstrated that the production
of the heavy $r$-elements must be associated with massive stars evolving on
short timescales, provided evidence strongly favoring core-collapse 
supernovae over neutron star mergers as 
the major source for these elements, and shown that this source 
cannot produce any significant amount of the low-$A$ elements from Na to Ge
including Fe. A self-consistent astrophysical model of the $r$-process 
remains to be developed, and it appears well worthwhile to carry out
more comprehensive studies on the evolution and explosion of the massive
stars of $\sim 8$--$11\,M_\odot$ that undergo O-Ne-Mg core collapse and
produce a very insignificant amount of the low-$A$ elements.}

\FullConference{10th Symposium on Nuclei in the Cosmos\\
                 July 27 - August 1, 2008\\
                 Mackinac Island, Michigan, USA}

\begin{document}

\section{Introduction}
This paper discusses some of the progress that has been made
in the understanding of the $r$-process since the publication of
\cite{cowa91}, a major review on this process, in 1991. More detailed discussion can be found in
\cite{qian03,qw07}. Readers are referred to \cite{arno07} for a review of the $r$-process that 
covers a wider range of topics and has different emphases.

For convenience of discussion, an $r$-process event can be separated into two phases. During
the first phase, relatively heavy nuclei are produced mostly via charged-particle reactions (CPRs).
These nuclei become the seed nuclei to capture the neutrons during the second phase. The crucial 
result of the first phase is a large (number) abundance ratio of the neutrons to the seed nuclei.
The second phase involves neutron-rich nuclei far from stability
participating in a number of types of nuclear reactions, which include neutron capture, 
photo-disintegration, $\beta$-decay, and possibly fission. The properties of such nuclei play
an essential role in determining the exact abundance pattern produced by the $r$-process event.
This important aspect of the $r$-process will not be discussed here. For the presentation below,
the abundance pattern resulting from the second phase is characterized by the
average mass number $\langle A_r\rangle$ of the corresponding $r$-process nuclei.
Assuming that all the neutrons are captured, the results from the first and 
second phases are related by mass conservation as
\begin{equation}
\langle A_s\rangle Y_s+Y_n=\langle A_r\rangle Y_r,
\label{eq-ns}
\end{equation}
where $\langle A_s\rangle$ is the average mass number of the seed nuclei, $Y_s$ is
their initial total (number) abundance, $Y_n$ is the initial neutron abundance, and
$Y_r$ is the final total abundance of the $r$-process nuclei produced.
In the absence of fission, the total number of heavy nuclei is conserved. For this case
$Y_r=Y_s$ and Equation~(\ref{eq-ns}) can be rewritten as
\begin{equation}
\langle A_s\rangle + \frac{Y_n}{Y_s}=\langle A_r\rangle,
\label{eq-ns2}
\end{equation}
which relates the average $r$-process nuclei to the average seed nuclei by the
neutron-to-seed ratio $Y_n/Y_s$.
Extensive fission cycling occurs for $Y_n/Y_s\gg\langle A_s\rangle$, in which case
Equation~(\ref{eq-ns}) approximately becomes
\begin{equation}
Y_n\approx\langle A_r\rangle Y_r.
\end{equation}
For this extreme case, a regular abundance pattern is expected from the steady-state
nature of the nuclear flow between the fission products and the fissioning nuclei, and
the final outcome is governed by the initial neutron abundance but not the exact 
abundance or form of the seed nuclei.

\begin{figure}
\vskip -0.75cm
\begin{center}
\includegraphics[angle=270,width=0.52\textwidth]{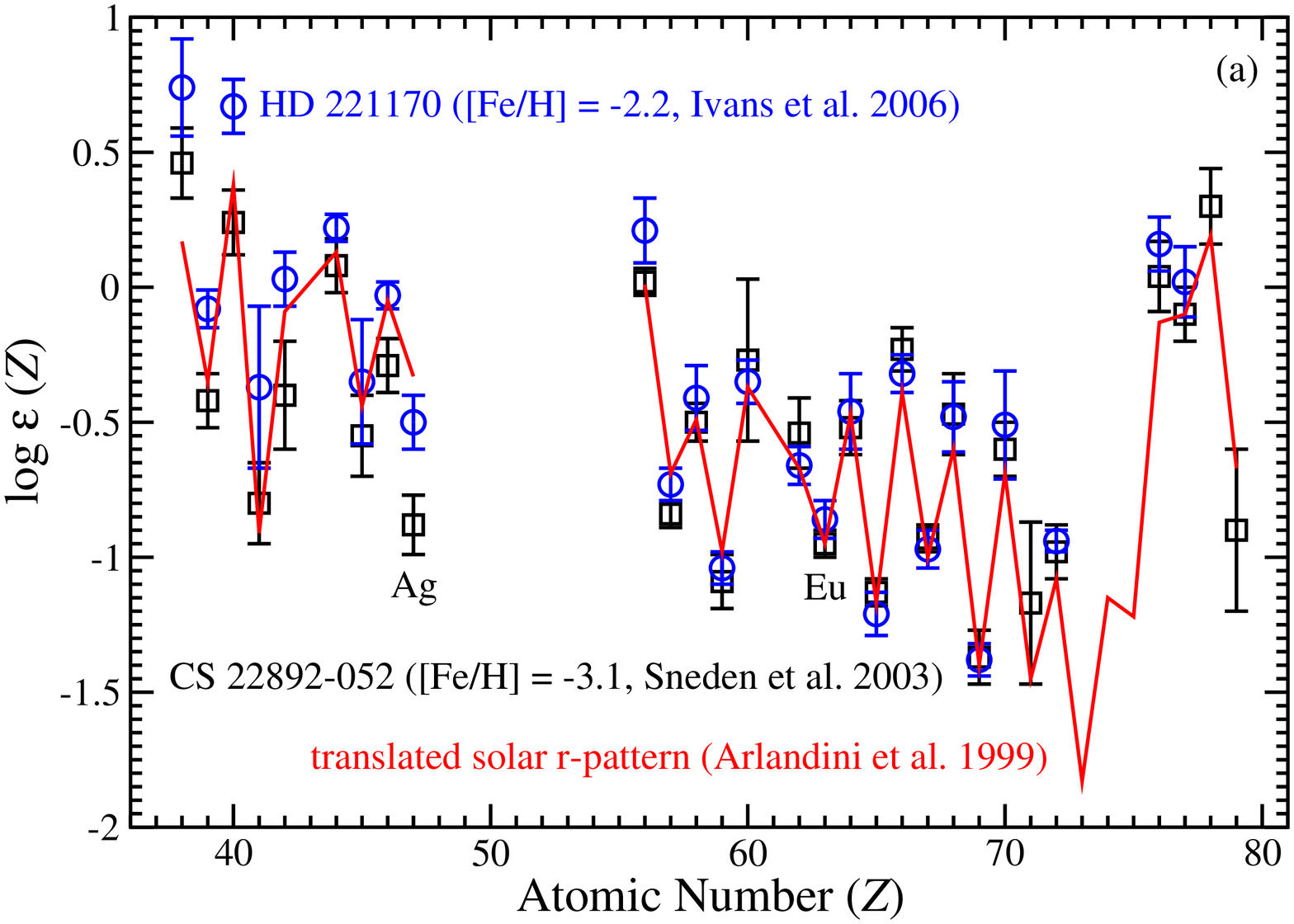}\hspace{-0.61cm}%
\includegraphics[angle=270,width=0.52\textwidth]{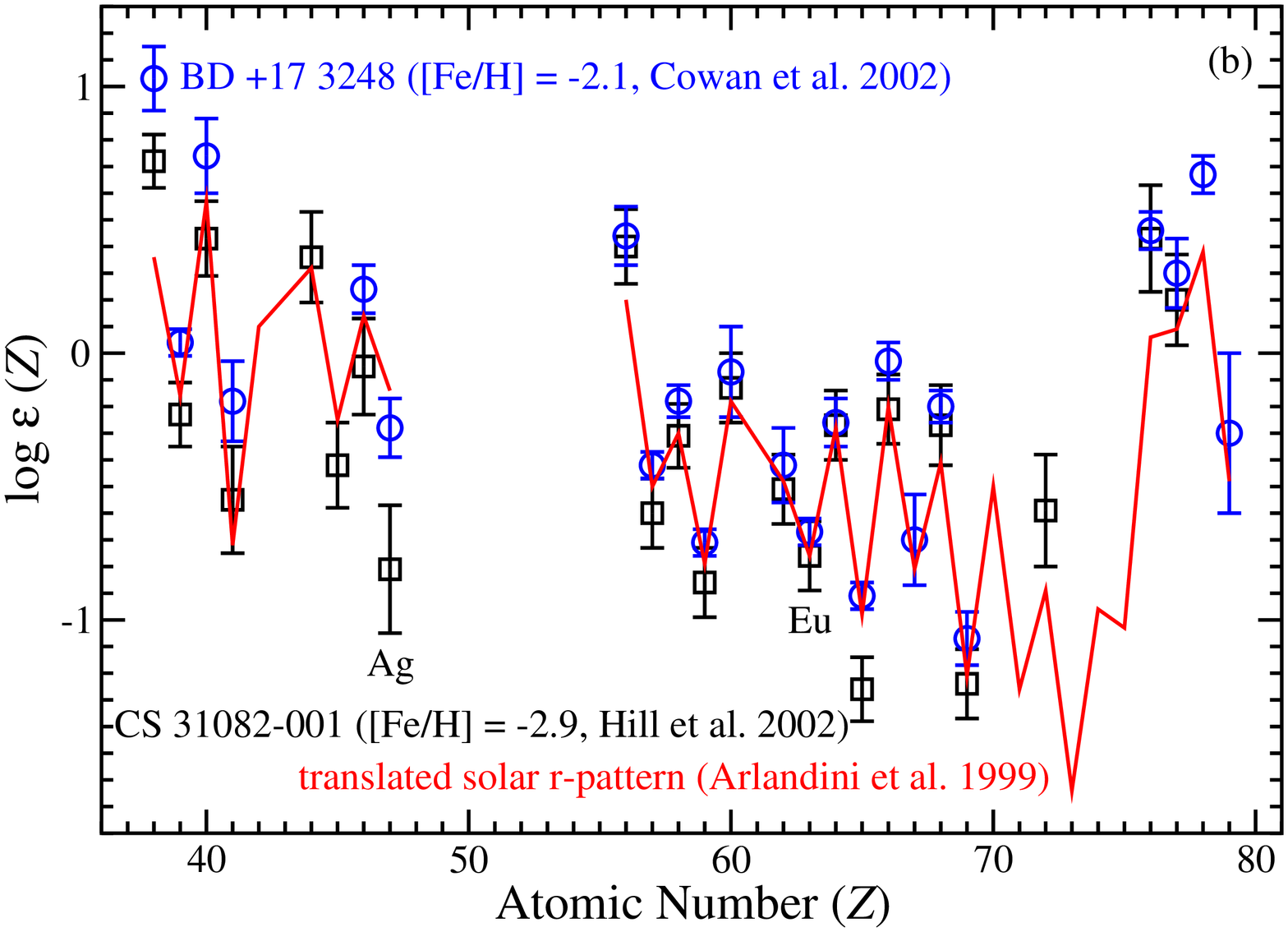}
\caption{Data on the $\log\epsilon$ values for the CPR elements from Sr to Ag
($Z=38$--47, $A\sim 88$--110) and the heavy $r$-elements from Ba to Au 
($Z=56$--79, $A>130$) for (a) CS~22892--052 (squares with error bars; \cite{sned03}) 
and HD~221170 (circles with error bars; \cite{ivan06}) and (b) CS~31082--001 
(squares with error bars; \cite{hill02}) and BD~$+17^\circ3248$ 
(circles with error bars; \cite{cowa02}). The solid curves in (a) and (b) represent
the solar ``$r$-pattern'' \cite{arla99} that is translated to pass through the Eu data for
CS~22892--052 and CS~31082--001, respectively. Note that the data on the heavy
$r$-elements follow the translated solar $r$-pattern rather well, especially for the
two stars in (a). However, the CPR elements, especially Sr and Ag, exhibit significant
to large deviations. As discussed in Section~\protect\ref{sec2}, these elements are not the 
true $r$-elements produced by rapid neutron capture.\label{fig1}}
\end{center}
\end{figure}

It can be seen from the above discussion that the neutron-to-seed ratio $Y_n/Y_s$ is
crucial to the success of an $r$-process event. Before discussing how this ratio is 
determined in the astrophysical models of these events, some general comments
regarding the $r$-process based on observations are in order. Figure~\ref{fig1} shows the
$\log\epsilon$ values\footnote{For an element E with a (number) abundance ratio (E/H) 
relative to hydrogen, $\log\epsilon({\rm E})\equiv\log({\rm E/H})+12$.} 
for the elements from Sr to
Ag with mass numbers $A\sim 88$--110 and those from Ba to Au with 
$A>130$ observed in four metal-poor stars CS~22892--052 \cite{sned03}, 
HD~221170 \cite{ivan06}, CS~31082--001 \cite{hill02}, and BD~$+17^\circ3248$
\cite{cowa02} with ${\rm [Fe/H]}\equiv\log{\rm (Fe/H)}-{\rm (Fe/H)}_\odot=-3.1$,
$-2.2$, $-2.9$, and $-2.1$, respectively. The solid curves in Fgiures~\ref{fig1}a
and \ref{fig1}b represent the so-called solar ``$r$-process'' abundance pattern 
($r$-pattern) that is translated to pass through the Eu data for CS~22892--052 and 
CS~31082--001, respectively. This pattern was derived by subtracting the 
$s$-process contributions calculated in \cite{arla99} from the total solar abundances. 
It can be seen from Figure~\ref{fig1} that the data for the elements with $A>130$ 
(Ba to Au with atomic numbers $Z=56$--79) 
in all four stars follow the translated solar $r$-pattern 
rather well, especially for the two stars in Figure~\ref{fig1}a. These elements will be
referred to as the heavy $r$-process elements ($r$-elements) hereafter.
However, the data for the elements with $A\sim 88$--110,
especially Sr and Ag ($Z=38$ and 47, respectively), 
exhibit significant to large deviations. As will be discussed in
Section~\ref{sec2}, these elements are routinely produced by the CPRs in the
neutrino-driven wind from a newly-formed neutron star, and therefore, are not the 
true $r$-elements produced by rapid neutron capture. They will be referred to as
the CPR elements hereafter. The remarkable agreement between the solar
$r$-pattern and the data for the heavy $r$-elements in four metal-poor stars
demonstrates that the $r$-process already occurred in the early Galaxy, and
therefore, must be associated with massive stars evolving on short
timescales. The rather regular abundance pattern exhibited by the heavy $r$-elements
also suggests that fission cycling may be involved in their production.

\section{Modeling and understanding the $r$-process}
\begin{figure}
\vskip 0.25cm
\begin{center}
\includegraphics[angle=270,width=0.6\textwidth]{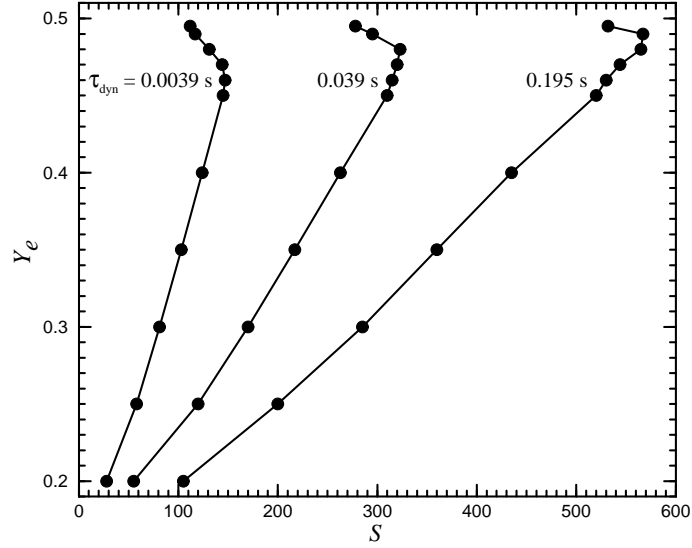}
\caption{Combinations of $Y_e$ and $S$ that would result in the major production of the heavy
$r$-elements with $A>130$ during the adiabatic expansion of matter from a hot and dense 
initial state for three values of $\tau_{\rm dyn}$. See \cite{hoff97} for details.\label{fig2}}
\end{center}
\end{figure}

A generic model of the $r$-process is associated with expansion of matter 
as it is ejected from a hot and dense initial state inside some astrophysical environment. 
At initial temperatures of $\gtrsim 1$~MeV, this matter is simply composed of free neutrons 
and protons with their abundances specified by the electron fraction $Y_e$ (matter is
neutron-rich for $Y_e<0.5$). The nucleosynthesis in this matter depends on 
the evolution of its temperature $T$ and density $\rho$ with time $t$ during the subsequent 
expansion. Assuming an adiabatic expansion with $T(t)\propto\exp(-t/\tau_{\rm dyn})$, 
where $\tau_{\rm dyn}$ is a dynamic timescale, the evolution of $\rho$ can be determined
from the conservation of entropy $S(T,\rho)$ (e.g., $S\propto T^3/\rho$ for 
radiation-dominated matter). In this case, the nucleosynthesis is governed by the three
parameters $Y_e$, $\tau_{\rm dyn}$, and $S$ of the expanding matter. The combinations
of $Y_e$ and $S$ that would result in the major production of the heavy $r$-elements with
$A>130$ for three values of $\tau_{\rm dyn}$ \cite{hoff97} (see also \cite{meye97,frei99})
are shown in Figure~\ref{fig2}.

The sets of $Y_e$, $S$, and $\tau_{\rm dyn}$ shown in Figure~\ref{fig2} correspond to
approximately the same neutron-to-see ratio $Y_n/Y_s$ that is required to produce the 
heavy $r$-elements according to Equation~(\ref{eq-ns2}). This ratio is determined as follows.
When the expanding neutron-rich matter cools to $T\sim 0.5$~MeV, essentially all the initial 
protons are assembled into $\alpha$-particles. An $\alpha$-process \cite{woho92} starts to 
burn the $\alpha$-particles and the remaining neutrons into heavier nuclei as $T$ further 
decreases. This process must go through the bottleneck of three-body reactions, mainly 
$\alpha+\alpha+n\to{^9{\rm Be}}+\gamma$. Once the bottleneck is passed, the subsequent
reactions starting with ${^9{\rm Be}}+\alpha\to{^{12}{\rm C}}+n$
proceed rather efficiently. The nuclear flow generally has reached the nuclei with 
$A\sim 90$ by the time all CPRs cease at $T\sim 0.25$~MeV due to the prohibitive Coulomb 
barrier. These nuclei then become the seed nuclei to capture the surviving neutrons during 
the ensuing $r$-process at lower $T$. Clearly, a lower $Y_e$ tends to give more neutrons,
i.e., a larger $Y_n$ at the end of the $\alpha$-process. A shorter $\tau_{\rm dyn}$ results in
fewer seed nuclei, i.e., a smaller $Y_s$ due to the shorter duration of the $\alpha$-process.
A higher $S$ also leads to a smaller $Y_s$. For the radiation-dominated matter with
$S\gg 10$ (in units of Boltzmann constant per baryon), there are many photons on the 
high-energy tail of the Bose-Einstein distribution that are able to incur the reaction
$\gamma+{^9{\rm Be}}\to\alpha+\alpha+n$, which requires a photon of only 1.573~MeV. 
Consequently, a high $S$ reduces the production of the seed nuclei by suppressing the
abundance of $^9$Be during the $\alpha$-process. In general, there are an infinite number
of sets of $Y_e$, $S$, and $\tau_{\rm dyn}$ that can result in the same $Y_n/Y_s$.
For example, Figure~\ref{fig2} shows that for a fixed $Y_e$, a lower $S$ can be 
combined with a shorter $\tau_{\rm dyn}$ to give the same $r$-process production.

\subsection{What is blown in the wind?\label{sec2}}
A specific model of $r$-process nucleosynthesis in adiabatically expanding matter is
associated with core-collapse supernovae (CCSNe) from massive stars. A few seconds
after the collapse of the core of a massive star, the shock ultimately responsible for 
disrupting the star has proceeded far into the envelope. The nascent neutron star
formed by the core collapse is cooling by emitting neutrinos. The matter immediately 
above the neutron star has $T>1$~MeV and is dissociated into free neutrons and protons.
As the neutrinos stream through this matter, a fraction of the $\nu_e$ and $\bar\nu_e$
are captured through the reactions $\nu_e+n\to p+e^-$ and $\bar\nu_e+p\to n+e^+$,
respectively. These reactions not only heat the matter and make it expand 
away from the neutron star but also interconvert neutrons and
protons, thereby determining the $Y_e$ of the matter \cite{qian93}. 
The neutrino heating becomes 
inefficient at distances of several neutron star radii, above which the matter expands 
essentially adiabatically in the form of a neutrino-driven wind with an approximately fixed
set of $Y_e$, $S$, and $\tau_{\rm dyn}$. It was proposed \cite{woos92} that 
the $r$-process occurs in such winds and this possibility has been extensively studied
(see e.g., \cite{meye92,taka94,woos94,wana01,wana06}). The physical conditions in 
the neutrino-driven winds from typical neutron stars have also been carefully studied
(see e.g., \cite{qian96,thom01}). Unfortunately, these conditions with typical values of
$Y_e\sim 0.4$, $S\sim 100$, and $\tau_{\rm dyn}\sim 0.1$~s (see also \cite{witt94}) 
do not match those
required for the major production of the heavy $r$-elements (see Figure~\ref{fig2}).

On the other hand, it was shown that the CPR elements from Sr to Ag with 
$A\sim 88$--110 are routinely produced by the $\alpha$-process in the 
neutrino-driven winds (see e.g., \cite{hoff97}). In fact, the production of these elements
was so overwhelming for the conditions obtained in some CCSN models that it was
regarded as a severe problem for such models (see e.g., \cite{meye92,woos94}).
However, this overproduction problem arises only for the winds with relatively
low $Y_e$ values that leave the neutron star at the earliest times. As the $Y_e$ 
of the wind is determined by the competition between $\nu_e$ capture on neutrons
and $\bar\nu_e$ capture on protons, it is extremely sensitive to the difference
between the rates of these reactions. For example, the $Y_e$ can be changed
substantially depending on whether the effects of the neutron-to-proton mass 
difference, nucleon recoil, and weak magnetism are included in the calculation of
these rates \cite{qian96,horo99}. These rates also depend on
the $\nu_e$ and $\bar\nu_e$ luminosities and energy spectra, which are taken
from CCSN neutrino transport calculations. It is conceivable that the relatively 
low $Y_e$ values of the earliest winds are caused by the uncertainties in
such calculations. As discussed in \cite{qw07,qw01}, observations of elemental 
abundances in metal-poor stars show that the CPR elements are not tightly 
coupled with the heavy $r$-elements. The production of the CPR elements 
in the neutrino-driven
winds is compatible with the non-production of the heavy $r$-elements therein.
The detailed chemical evolution of the CPR elements in the early Galaxy is
discussed in \cite{qw08,wass08} based on their association with neutron
star formation.

\subsection{Where, oh where has the $r$-process gone?\label{sec22}}
The question that remains to be answered is where the $r$-process occurs.
Some workers have explored the effects of e.g., the neutron star magnetic field 
\cite{thom03} on the neutrino-driven winds in the hope of obtaining the conditions 
required for the major production of the heavy $r$-elements. Others have 
proposed different sites for the $r$-process within the general context of CCSNe,
such as winds or jets from accretion disks of black holes 
(see e.g., \cite{came01,prue03,mcla05}) or the material undergoing complicated evolution
from initial fall-back to final ejection \cite{frye06}. These studies are focused on
obtaining the conditions required for the $r$-process. The approach here is different
and uses observations of elemental abundances in stars to constrain
the $r$-process site based on considerations of Galactic chemical evolution and
a basic understanding of stellar physics. 
This approach can be illustrated by contrasting CCSNe against neutron
star mergers (NSMs; see e.g., \cite{frt99,ross99}) as the major 
source for the heavy $r$-elements (see e.g., \cite{qian00,arga04}).

\begin{figure}
\begin{center}
\includegraphics[width=0.97\textwidth]{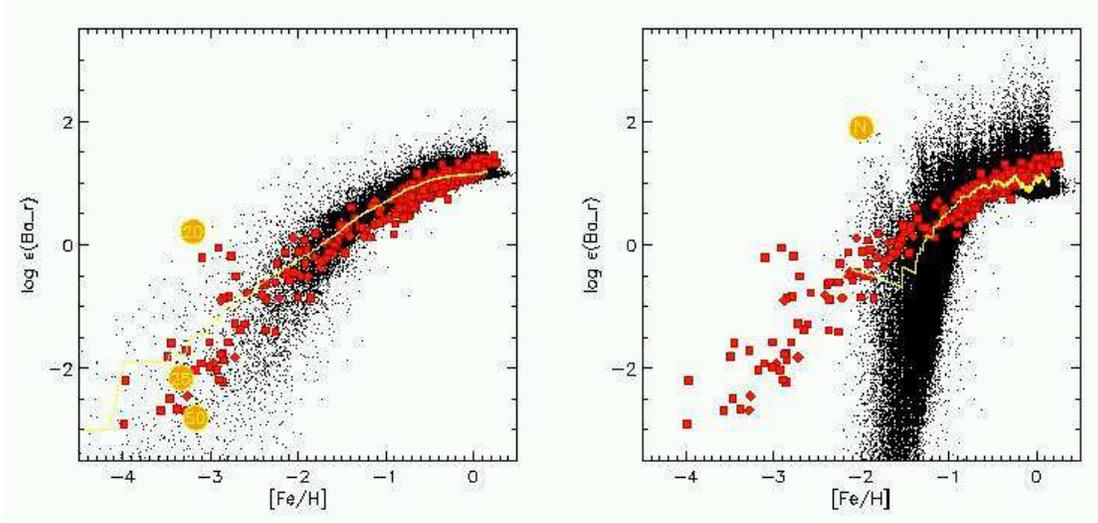}
\caption{Resulting distribution (dots) of $\log\epsilon({\rm Ba})_r$ vs. [Fe/H] 
if CCSNe (left panel) or NSMs (right panel) are the major source for the heavy 
$r$-elements. The data on stars are shown as filled squares. It can be seen
that the data strongly disfavor NSMs as the major source
for the heavy $r$-elements. See \cite{arga04} for details.\label{fig3}}
\end{center}
\end{figure}

The dots in Figure~\ref{fig3} represent the resulting distribution of 
$\log\epsilon({\rm Ba})_r$ vs. [Fe/H] if CCSNe (left panel) or NSMs (right panel) are 
the major source for the heavy $r$-elements \cite{arga04}. Here 
$\log\epsilon({\rm Ba})_r$ corresponds to the Ba contributed by the $r$-process.
As NSMs do not produce Fe,
the sources for Fe are the same for both cases. CCSNe are the only Fe source
at [Fe/H]~$\lesssim -1.5$, while both CCSNe and Type Ia SNe contribute Fe
at [Fe/H]~$>-1.5$. Figure~\ref{fig3} shows that there is a sudden rise of
$\log\epsilon({\rm Ba})_r$ at [Fe/H]~$\sim -2$ and wide dispersion in
$\log\epsilon({\rm Ba})_r$ at any [Fe/H] value for [Fe/H]~$\gtrsim -2$
if NSMs are the major source for the heavy $r$-elements. This can be
understood from the difference in the rate of occurrences between CCSNe
and NSMs. The Galactic rates of CCSNe and NSMs are $\sim 10^{-2}$ and
$\sim 10^{-5}$~yr$^{-1}$, respectively. If NSMs were the major source for
the heavy $r$-elements, many CCSNe would have enriched an interstellar
medium (ISM) with Fe before an NSM provided Ba to this ISM, which
explains the sudden rise of $\log\epsilon({\rm Ba})_r$ at the relatively
high value of ${\rm [Fe/H]}\sim -2$ (right panel of Figure~\ref{fig3}).
Further, the low Galactic rate of NSMs means that very few NSMs could
have occurred in an average ISM over the entire Galactic history \cite{qian00},
which explains the wide dispersion at any [Fe/H] value for [Fe/H]~$\gtrsim -2$
(right panel of Figure~\ref{fig3}). The filled squares in Figure~\ref{fig3} show the
data on $\log\epsilon({\rm Ba})_r$ vs. [Fe/H] for stars. It can be seen that
the data strongly disfavor NSMs as the major source for the heavy $r$-elements.
In contrast, a major CCSN source for these elements is compatible with the data.

\begin{figure}
\begin{center}
\includegraphics[width=0.46\textwidth]{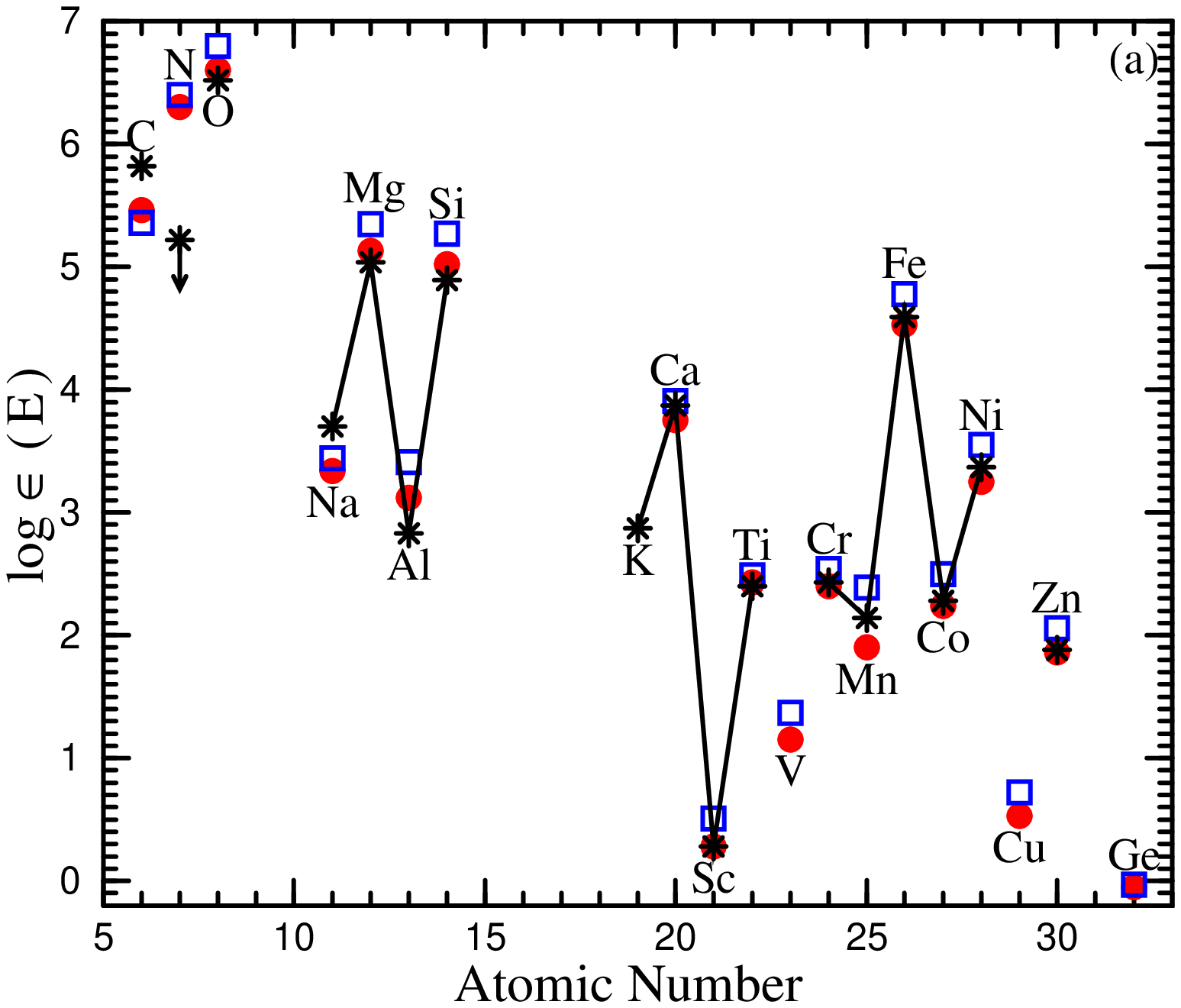}\hskip 0.45cm%
\includegraphics[width=0.5\textwidth]{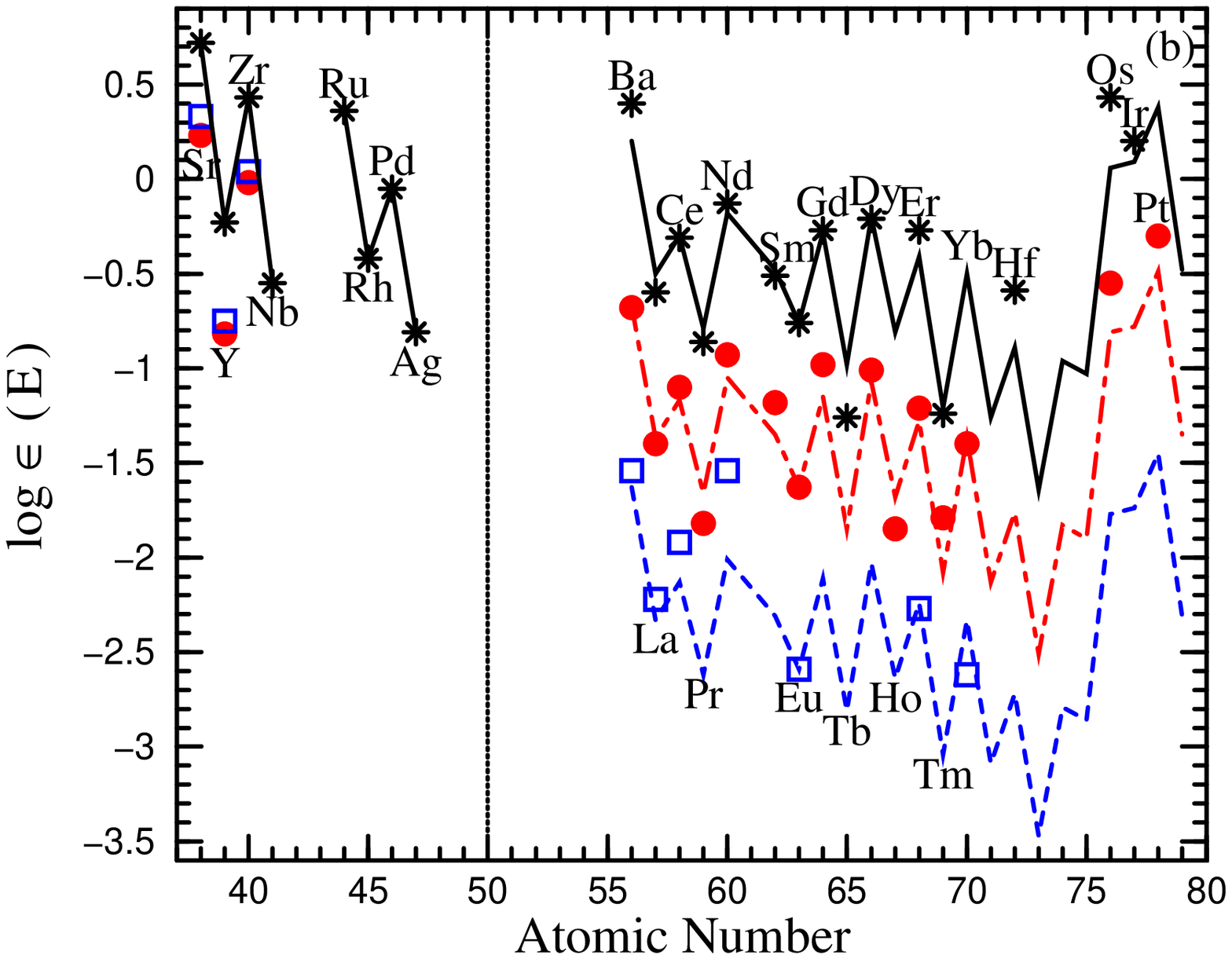}
\caption{Data on the elements from C to Pt in CS~31082--001 (asterisks;
\cite{hill02}), HD~115444 (filled circles), and HD~122563 (squares; \cite{west00})
with [Fe/H]~$=-2.9$, $-2.99$, and $-2.74$, respectively.
(a) The $\log\epsilon$ values for the elements from C to Ge. The data
on CS~31082--001 are connected by solid line segments as a guide.
The downward arrow at the asterisk for N indicates an upper limit. 
Note that the available abundances for the elements from O to Ge are almost 
indistinguishable for the three stars. (b) The $\log\epsilon$ values for 
the elements from Sr to Pt. The data for CS~31082--001 in the region to the left of 
the vertical dotted line are again connected by solid line segments as a guide. 
In the region to the right of the vertical dotted line, the data on
the heavy $r$-elements are compared with the solid, dot-dashed, and 
dashed curves, which are the solar $r$-pattern \cite{arla99} translated to
pass through the Eu data for CS~31082--001, HD~115444, and
HD~122563, respectively. Note the general agreement between the
data and the solid and dot-dashed curves. There is a range of 
$\sim 2$~dex in the abundances of the heavy $r$-elements for the
three stars shown.\label{fig4}}
\end{center}
\end{figure}

The above approach also provides important information on the specific characteristics
of the major source for the heavy $r$-elements. Figure~\ref{fig4} shows the 
$\log\epsilon$ values for the elements from C to Pt in three metal-poor stars
CS~31082--001 \cite{hill02}, HD~115444, and HD~122563 \cite{west00} with 
[Fe/H]~$=-2.9$, $-2.99$, and $-2.74$, respectively. The abundances of the heavy
$r$-elements in these stars differ by a factor of $\sim 100$ (see Figure~\ref{fig4}b).
However, they have essentially the same abundances of the low-$A$ elements 
from Na to Ge (including Fe)
with $A\sim 23$--70 (see Figure~\ref{fig4}a). This shows that the 
production of the heavy $r$-elements must be completely decoupled from that of
the low-$A$ elements \cite{qw02}. This decoupling is also demonstrated by the comparison
of CS~22892--052 ([Fe/H]~$=-3.1$) with HD~221170 ([Fe/H]~$-2.2$) and that of 
CS~31082--001 ([Fe/H]~$=-2.9$) with BD~$+17^\circ3248$ ([Fe/H]~$=-2.1$)
in Figures~\ref{fig1}a and \ref{fig1}b, respectively. The stars in each pair have 
essentially the same abundances of the heavy $r$-elements. However,
as indicated by the Fe abundances, the abundances of the low-$A$ 
elements\footnote{Observations (see e.g., \cite{cayr04}) show that the abundance 
patterns of the low-$A$ elements are rather uniform for nearly all the metal-poor stars
with $-4\lesssim{\rm [Fe/H]}\lesssim -1.5$.
This is true for the stars shown in Figure~\ref{fig1} in particular. So the Fe abundance
reflects the level of enrichment for all the low-$A$ elements in these stars.} 
differ by factors of $\sim 8$ and $\sim 6$ for the paired stars shown in 
Figures~\ref{fig1}a and \ref{fig1}b, respectively. 

The observations discussed above require that the major source for the
heavy $r$-elements produce none of the low-$A$ elements. This means that
normal CCSNe from massive stars of $\gtrsim 12\,M_\odot$ cannot be such a
source. These stars develop an Fe core surrounded by extensive 
shells at the end of their lives and produce the low-$A$ elements by the
hydrostatic burning in the outer shells during the pre-SN evolution and
by the explosive burning associated with the propagation of the SN
shock through the inner shells (see e.g., \cite{woos95}). In contrast,
massive stars of $\sim 8$--$11\,M_\odot$ develop an O-Ne-Mg core
with thin shells of very little mass between the core and the hydrogen 
envelope (see e.g., \cite{nomo84,nomo87}). Consequently, the CCSNe 
from these stars produce a very insignificant amount of the low-$A$ 
elements, thereby satisfying the observational requirement on the major 
source for the heavy $r$-elements. This has led to the proposal that
these low-mass CCSNe are this major source \cite{qw03}. A subsequent
work further proposed that the heavy $r$-elements are produced as the
SN shock rapidly accelerates through the surface layers of an
O-Ne-Mg core \cite{ning07}. However, the conditions adopted by this
work are not readily obtained in the current SN models \cite{jank08} 
based on the pre-SN structure calculated in \cite{nomo84,nomo87}.

\section{Conclusions}
There has been significant progress in the understanding of the $r$-process
through both theoretical and observational studies over the last few decades.
The conditions required for the major production of the heavy $r$-elements 
have been derived for a generic astrophysical model where
matter adiabatically expands from a hot and dense initial state. Nucleosynthesis
in the neutrino-driven winds from nascent neutron stars has been studied
extensively. While such winds readily produce the elements from Sr to Ag 
through charged-particle reactions in the $\alpha$-process, they 
appear incapable of making the heavy $r$-elements.
The insights provided by observations of elemental abundances in metal-poor
stars are especially important. These observations
have demonstrated that the production
of the heavy $r$-elements must be associated with massive stars evolving on
short timescales, provided evidence strongly favoring CCSNe over NSMs as 
the major source for these elements, and shown that this source cannot
produce any significant amount of the low-$A$ elements from Na to Ge
including Fe. A self-consistent astrophysical model of the $r$-process 
remains to be developed, and it appears well worthwhile to carry out
more comprehensive studies on the evolution and explosion of the massive
stars of $\sim 8$--$11\,M_\odot$ that undergo O-Ne-Mg core collapse and
produce a very insignificant amount of the low-$A$ elements.

\end{document}